# Raman spectroscopy and X-ray diffraction of $sp^3$-CaCO$_3$ at lower mantle pressures

Sergey S. Lobanov[1,2], Xiao Dong[3], Naira S. Martirosyan[1,2], Artem I. Samtsevich[4], Vladan Stevanovic[5], Pavel N. Gavryushkin[2,6], Konstantin D. Litasov[2,6], Eran Greenberg[7], Vitali B. Prakapenka[7], Artem R. Oganov,[4,8,9,10], and Alexander F. Goncharov[1,11]

[1]Geophysical Laboratory, Carnegie Institution of Washington, Washington, DC 20015, USA

[2]Sobolev Institute of Geology and Mineralogy Siberian Branch Russian Academy of Sciences, 3 Pr. Ac. Koptyga, Novosibirsk 630090, Russia

[3]Center for High Pressure Science and Technology Advanced Research, Beijing 100193, China

[4]Skolkovo Institute of Science and Technology, Skolkovo Innovation Center, 5 Nobel St., Moscow 143026, Russia

[5]Department of Metallurgical and Materials Engineering, Colorado School of Mines, Golden, CO 80401, USA

[6]Novosibirsk State University, Novosibirsk 630090, Russian Federation

[7]Center for Advanced Radiations Sources, University of Chicago, Chicago, IL 60632, USA

[8]Moscow Institute of Physics and Technology, 9 Institutskiy Lane, Dolgoprudny City, Moscow Region 141700, Russia

[9]School of Materials Science, Northwestern Polytechnical University, Xi'an 710072, China

[10]Department of Geosciences, Center for Materials by Design, Institute for Advanced Computational Science, Stony Brook University, Stony Brook, New York 11794, United States

[11]Key Laboratory of Materials Physics, Institute of Solid State Physics CAS, Hefei 230031, China

*E-mail: slobanov@carnegiescience.edu; slobanov@igm.nsc.ru

**Abstract**

The exceptional ability of carbon to form $sp^2$ and $sp^3$ bonding states leads to a great structural and chemical diversity of carbon-bearing phases at non-ambient conditions. Here we use laser-heated diamond anvil cells combined with synchrotron x-ray diffraction, Raman spectroscopy, and first-principles calculations to explore phase transitions in CaCO$_3$ at P > 40 GPa. We find that post-aragonite CaCO$_3$ transforms to the previously predicted $P2_1/c$-CaCO$_3$ with $sp^3$-hybridized carbon at 105 GPa (~30 GPa higher than the theoretically predicted crossover pressure). The lowest enthalpy transition path to $P2_1/c$-CaCO$_3$ includes reoccurring $sp^2$- and $sp^3$-CaCO$_3$ intermediate phases and transition states, as reveled by our variable-cell nudged elastic band simulation. Raman spectra of $P2_1/c$-CaCO$_3$ show an intense band at 1025

cm$^{-1}$, which we assign to the symmetric C-O stretching vibration based on empirical and first principles calculations. This Raman band has a frequency that is ~20 % lower than the symmetric C-O stretching in $sp^2$-CaCO$_3$, due to the C-O bond length increase across the $sp^2$-$sp^3$ transition, and can be used as a fingerprint of tetrahedrally-coordinated carbon in other carbonates.

**Introduction**

The thermodynamic ground state of carbon at ambient conditions is graphite with a triangular bonding pattern ($sp^2$ hybridization). High pressure (P), however, favors tetrahedrally-bonded ($sp^3$) carbon, and diamond is stable at P > 1.7 GPa (0 K) [1]. The different bonding patterns of graphite and diamond result in very different mechanical, optical, electric, and thermal properties [2], making carbon a truly remarkable element. On top of this, the binding energy between carbon atoms is very large leading to high melting temperatures (T) as well as high activation energies for the solid state phase transitions [1]. As a result, carbon has a rich variety of metastable phases with mixed $sp^2$ and $sp^3$ bonding patterns that may integrate the unique physical properties of both graphite and diamond [3,4]. The synthesis of such novel carbon-based technological materials requires navigating in the carbon energy landscape as well as insights into the trajectories and mechanisms of its phase transitions [5].

Unlike carbon, the thermodynamically stable form of silicon at ambient condition has the cubic diamond structure. Not surprisingly, nearly all low-pressure silicates incorporate silicon exclusively in the form of $sp^3$-hybridized SiO$_4$ tetrahedral groups. The electronic structure of SiO$_4$-tetrahedra is such that each oxygen has a half-occupied $p$ orbital available for polymerization with adjacent groups. The topology of polymerized SiO$_4$-networks largely governs the physical properties of silicates and serves as the basis for their structural classification [6,7]. On the other hand, $sp^2$-hybridized CO$_3$ triangular groups have an additional C-O π bond, and as a result, are isolated in the crystal structures of carbonates. This difference in the electronic structures of CO$_3$ and SiO$_4$ groups leads to very different physical properties of $sp^2$-carbonates and $sp^3$-silicates. At high pressure, however, the electronic structure of carbon in carbonates may change via the C-O π bond polymerization as individual CO$_3$ groups approach each other. Theoretical computations predict that $sp^3$-carbonates become thermodynamically stable at P > ~ 80-130 GPa [8-12]. Here we investigate the high-P behavior of CaCO$_3$, one of the most abundant carbonates near the Earth's surface and a good proxy for carbonate chemical composition in the mantle [13,14].

Previous high-P studies have revealed a number of pressure-induced transformations in $CaCO_3$. At P < ~ 40 GPa, (meta)stable phases of $CaCO_3$ include calcite, aragonite, $CaCO_3$-II, $CaCO_3$-III, CaCO3-IIIb, and $CaCO_3$-VI (*e.g.* [9,15-17]). At P > 40 GPa, $CaCO_3$ transforms into post-aragonite, which has been reported as a stable phase up to 137 GPa [9,18,19]. Importantly, all these structures contain $sp^2$-hybridized carbon forming triangular $CO_3$ groups. Pyroxene-like $C222_1$-$CaCO_3$, which has been predicted stable at P > 137 GPa, has a different bonding pattern with $sp^3$-hybridized carbon forming polymerized $CO_4$-chains [9]. This prediction gained some experimental support in that the major Bragg peaks of the $C222_1$-$CaCO_3$ had been observed in experiment at P > 140 GPa [19]. The high synthesis pressure implied that $sp^3$-$CaCO_3$ is not present in the Earth's mantle (135 GPa is the core-mantle boundary pressure) and further experimental studies of $sp^3$-carbonates were shifted to other compositions. More recently, the $sp^2$-$sp^3$ transition in $CaCO_3$ was revisited by Pickard and Needs [12] who predicted a new $sp^3$-$CaCO_3$ phase ($P2_1/c$) at P > 76 GPa calling for a new synthesis study.

Here we explore phase transitions in $CaCO_3$ at P > 40 GPa via synchrotron x-ray diffraction, Raman spectroscopy, and first-principles calculations. We establish the stability field of $sp^3$-bonded $P2_1/c$-$CaCO_3$ and show that this phase has a strong Raman band characteristic of fourfold carbon in its crystal structure. We provide computational insights into the $sp^2$-$sp^3$ phase transition mechanism, which in $CaCO_3$ appears to be a complex multistage process. Finally, our results support the notion of the effect of $sp^2$-$sp^3$ crossover on the carbonates crystal chemistry in the lower mantle.

**Methods**

**Experimental methods.** Diamond anvil cells (DACs) equipped with flat 200-300 μm culets were used to generate high pressure. Rhenium gaskets (~ 200 μm thick) were indented to ~ 30-40 μm by the anvils and laser-drilled in the center of the indentation in order to prepare a sample chamber with a diameter of 70-120 μm. The sample chamber was loaded with 99.95 % $CaCO_3$ (Alfa Aesar) mixed with Pt powder (20-30 %) which served both as a heating laser absorber and as a pressure standard [20]. No pressure-transmitting medium was used in the experiments.

X-ray diffraction (XRD) measurements and laser-heatings were performed at the 13ID-D GeoSoilEnviroCARS beamline (Argonne National Lab, APS) that allows *in situ* XRD collections at extreme P-T conditions and a subsequent high resolution mapping of the sample quenched to ambient temperature [21]. At all pressures a typical heating cycle involved: (i) heating to T ~ 2000 K, while following the diffraction pattern each 100-200 K; (ii) annealing at

T ~ 2000 K, at which temperature we typically observed the formation of new XRD peaks, while moving the samples by ~ 10 μm in horizontal and vertical directions (1 μm step); (iii) quenching and mapping the heated region in order to find areas with less Pt and more $CaCO_3$. The x-ray energy was 37-42 keV focused to ~ 3 by 4 μm spot. 2D XRD images were integrated using the Dioptas software [22] for on-line analyses. Selected XRD patterns were analyzed in PowderCell 2.4 and LeBail-refined in GSAS/EXPGUI [23,24]. Equation of state fitting (EOS) was performed using EoSFit7GUI [25] and VESTA [26] for structure visualization.

After the synthesis and XRD measurements, samples with $sp^3$-$CaCO_3$ were characterized by Raman spectroscopy upon decompression at the Geophysical Laboratory using solid-state 488 (Spectra-Physics), 532 (Laser Quantum GEM), and 660 nm (Laser Quantum Ignis) laser-excitations focused to a 3-4 μm spot size in diameter. The use of three excitation wavelengths allows unambiguously identifying bands that are Raman in origin. Backscattered Raman radiation was spatially filtered through a 50 μm pinhole (magnified by 10 using a Mitutoyo 20X NA0.4 long working length objective lens) to eliminate spurious signal and collected by custom Raman spectrometers with CCD array detectors (PIXIS 100, Princeton Instruments) equipped with same-turret 300 and 1200/1500 grooves per mm gratings (HR 460, JOBIN YVON for the 488 nm setup and Acton SP2300/2500 of Princeton Instruments for 532 nm 660 nm, respectively). The spectral resolution was ~ 4 $cm^{-1}$. The diamond Raman edge stress scale [27] was used to determine pressure on decompression with an uncertainty of ~ 3-5 GPa.

**Theoretical methods**

In this study we relied on the previous structural searches [12] but the use of USPEX yields similar results (not presented here). Structural relaxations and Raman intensity calculations were performed based on the density functional theory (DFT) as implemented in the Quantum-ESPRESSO code [28]. The norm conserving pseudopotential [29] was used and the electron-electron exchange and correlation was described by the local density approximation (LDA) exchange-correlation functional of Ceperley and Alder, as parameterized by Perdew and Zunger (CA-PZ) [30]. The plane-wave cutoff energy with 250 Ry, and a k-point spacing ($2\pi \times 0.03$ $Å^{-1}$) was used to generate Monkhorst-Pack k-points grids for Brillouin zone sampling [31].

**Results and Discussion**

**X-ray diffraction.** Room-temperature compression to P > 40 GPa results in a diffraction pattern with several low intensity diffuse peaks. Annealing the samples at 40-102 GPa and 1500-2000 K produces new sharp Bragg reflections that can be indexed with the post-aragonite (*Pmmn*) $CaCO_3$ phase [9,18] (Supplementary Fig. S1). At 105 GPa, the dominant annealing product is different and forms a new spotty pattern in the XRD images (Fig. 1), but residual broad and diffuse reflections of precursor CaCO$_3$ are also present after the heating. Crystallographic indexing of the new reflections yields monoclinic and orthorhombic solutions with unit cells consistent with the theoretical predictions of $C222_1$ [11] and $P2_1/c$ [12] CaCO$_3$.

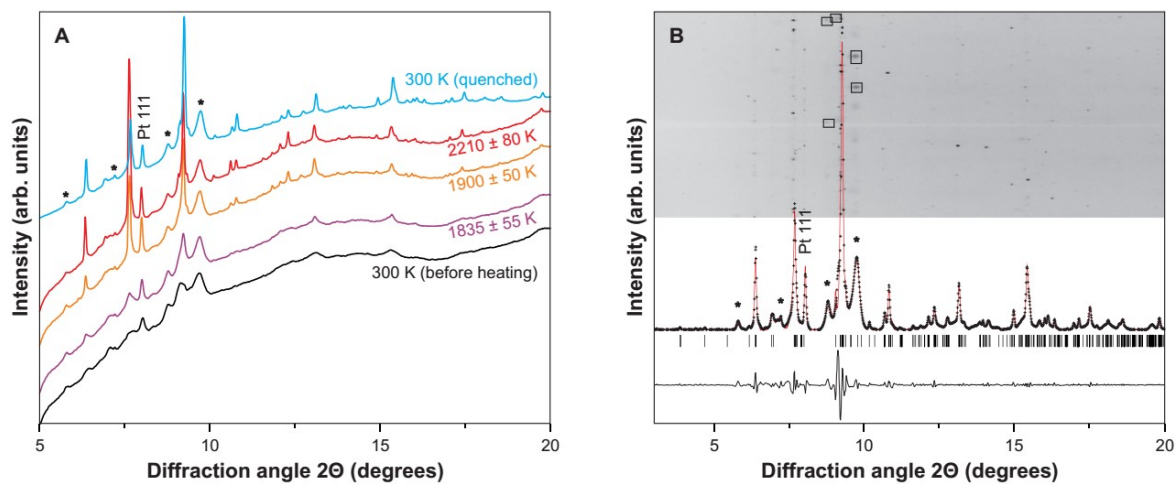

**Figure 1.** (**A**) X-ray diffraction (XRD) of CaCO$_3$ before, at *T*, and after heating at 105 GPa (with background). (**B**) LeBail fit of the theoretically predicted $P2_1/c$-CaCO$_3$ (red line) to the experimentally observed XRD pattern (black crosses). Thin black line is the difference curve. The corresponding rectangular diffraction image is shown in the upper part of (**B**). Asterisks and black boxes mark some of the diffuse peaks of remnant CaCO$_3$. X-ray energy is 42 keV.

Both theoretically proposed models allow indexing the new peaks yielding almost identical densities at 105 GPa (5.01(2) g/cm$^3$). Indeed, topological analysis, performed to reveal structural differences between the two $sp^3$-CaCO$_3$ structures, shows a high degree of similarity between the $P2_1/c$ and $C222_1$ structures with an identical atomic coordination (Ca$^{[10]}$C$^{[4]}$O$_2^{(5)}$O$^{(4)}$) and arrangement of Ca and C atoms. The only difference between the structures is the orientation of CO$_4$-tetrahedra: all vertex-sharing helices in $C222_1$-CaCO$_3$ are right-handed, while half helices in the $P2_1/c$-CaCO$_3$ are left-handed (Fig. 2). Despite of these similarities, the $P2_1/c$ structure has an approximately 0.2 eV/f.u. lower enthalpy than $C222_1$-CaCO$_3$, according to the computation of Pickard and Needs [12], advocating in favor of the monoclinic structure. Here we provide further support for the $P2_1/c$-CaCO$_3$ as its structural model allows indexing severely split peaks, such as the -112 and 111 Bragg reflections at ~ 7 degrees and the feature at ~ 9.2 degrees 2Θ, as well as other minor reflections in the observed XRD pattern (Fig. 1B). Accordingly, LeBail refinements of the XRD patterns with the $P2_1/c$ structure systematically yield ~ 5 % better fits than that performed with the $C222_1$ structure. Please note that although we could not perform a full-profile refinement in this work due to the textured XRD pattern, the observed intensities are also consistent with the $P2_1/c$ model (Supplementary Fig. S2). Hence, we confirm the prediction of the $P2_1/c$-CaCO$_3$, albeit at ~ 30 GPa higher than the theoretically predicted $sp^2$-$sp^3$ crossover pressure [12]. We note that although $P2_1/c$ and $C222_1$ models of CaCO$_3$ have very similar powder XRD patterns, their Raman spectra may bear significant differences and may help to identify the $sp^3$-CaCO$_3$ phase.

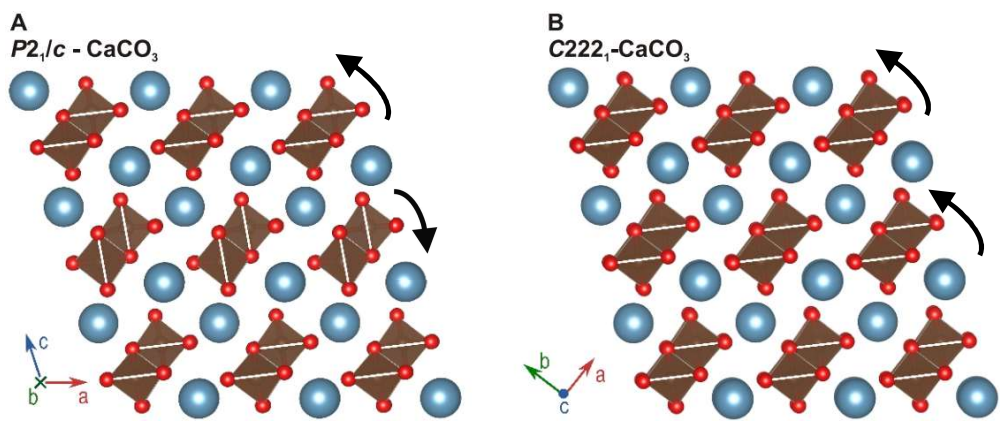

**Figure 2.** Structures of $P2_1/c$-CaCO$_3$ (**A**) and $C222_1$-CaCO$_3$ (**B**) with outlined CO$_4$-tetrahedra. Calcium atoms are shown in blue, carbon in brown, and oxygen in red. Black arrows show the distinct chirality of CO$_4$-tetrahedra chains in the crystal structures.

Depending on the probed sample area, we observed a coexistence of the post-aragonite phase with $P2_1/c$-CaCO$_3$ at 103-105 GPa, which indicates that this pressure is close to the phase transition pressure. At 105 GPa and 300 K, the unit cell parameters of post-aragonite CaCO$_3$ are $a = 3.9360(6)$ Å, $b = 4.4372(3)$ Å, $c = 3.9049(4)$ Å ($\rho = 4.87(2)$ g/cm$^3$), while that of $P2_1/c$-CaCO$_3$ are $a = 4.5288(13)$ Å, $b = 3.3345(3)$ Å, $c = 9.0927(24)$ Å, and $\beta = 105.57(9)$ degrees ($\rho = 5.01(2)$ g/cm$^3$) (Supplementary Table S1). The structure of $sp^3$-CaCO$_3$ is ~ 3 % denser than that of its $sp^2$-bonded counterpart at 105 GPa (Fig. 3), which is larger than the previously reported density contrasts of 0.5 % [19] and 1.25 % [9] across the $sp^2$-$sp^3$ transition. Importantly, the average carbon-oxygen bond length increases across the phase transition from 1.228 to 1.315 Å (by ~ 7 %) as a result of the increased carbon coordination. Note that in order to determine the change in C-O bond length over the $sp^2$-$sp^3$ transition in CaCO$_3$ we used the experimentally refined lattice parameters of the coexisting CaCO$_3$ phases at 105 GPa and theoretically computed atomic positions [12]. Although we did not refine the atomic positions based on the experimental XRD, the observed intensities are consistent with the theoretically-proposed $P2_1/c$-CaCO$_3$ model (Supplementary Fig. S2). Because of the increase in C-O bond length, one would expect an abrupt decrease in the frequency of the carbon-oxygen stretching vibration across the $sp^2$-$sp^3$ transition.

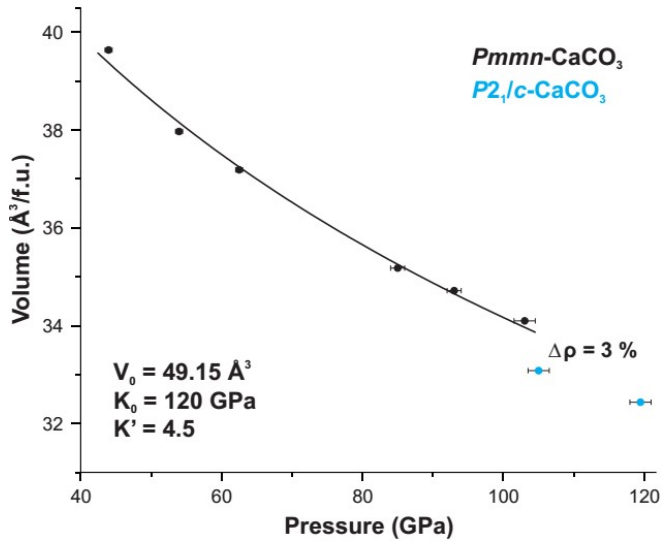

**Figure 3.** Pressure-volume relations for *Pmmn*-CaCO$_3$ (black dots) and *P*2$_1$/*c*-CaCO$_3$ (blue dots). Black line is a 300 K third-order Birch-Murnaghan equation of state (EOS) of *Pmmn*-CaCO$_3$ (post-aragonite) fitted to the collected here P-V data. Best fits were obtained using the previously reported post-aragonite V$_0$ value (49.15 Å$^3$/f.u.) [18] in combination with *K'* in the range of 4.5-4.7. Fixing V$_0$ to the reported value is appropriate because of the larger number of P-V measurements in the previous study. Corresponding EOS parameters are given in the bottom left corner. Pressure uncertainty (σ) is assumed to be 0.5, 1, and 1.5 GPa for P < 70, 80-100, and > 100 GPa, respectively.

**Raman spectroscopy.** Group theory for *P*2$_1$/*c*-CaCO$_3$ allows 30 Raman active vibrations (15A$_g$ + 15B$_g$). Raman spectra collected from the laser-heated area consistently show at least 8 new peaks all of which appear characteristic of the vibrational normal modes in the new carbonate as the frequency and relative intensity of these bands are independent of the excitation wavelength (Fig. 4). Particularly important is the new intense band at 1025 cm$^{-1}$. Considering the increased C-O bond length across the *sp$^2$*-*sp$^3$* transition it is reasonable to suppose that this high-frequency band corresponds to the C-O stretching vibration in the CO$_4$-unit. We have a rough check on this assignment by assuming a harmonic oscillator and an empirically established relation of the force constant and bond length for CX compounds [32]: $f = a(r - 0.61)^{-3}$, where X is a second period element, *a* is a constant, and *r* is the C-X equilibrium bond length. Accepting the change in C-O bond length across the *sp$^2$*-*sp$^3$* transition as well as the frequency of C-O symmetric stretching vibration in *sp$^2$*-CaCO$_3$ at 105 GPa (1290 cm$^{-1}$) we obtain a frequency of 1059 cm$^{-1}$ for this vibration in *sp$^3$*-CaCO$_3$. This is within 5 % with the observed frequency of 1025 cm$^{-1}$ in support of its assignment to the C-O symmetric stretching in tetrahedral-coordinated carbon. A similar comparison for the graphite-diamond C-C stretch modes yields a frequency of 1273 cm$^{-1}$ for the diamond T$_{2g}$ band at 1 atm, which is again < 5 % off its actual value (1333 cm$^{-1}$).

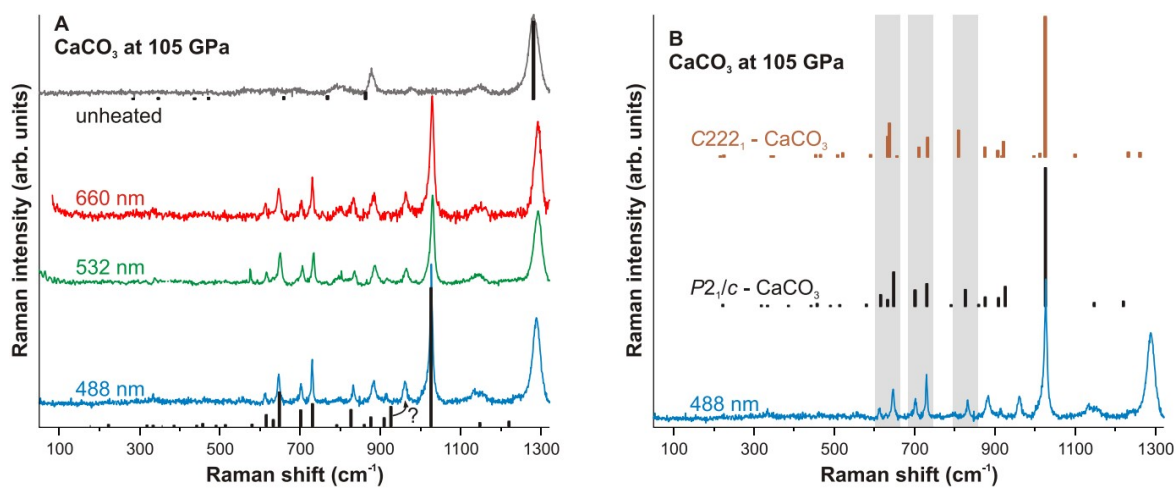

**Figure 4.** (**A**) Raman spectra of CaCO$_3$ at 105 GPa collected with 488, 532, and 660 nm excitations. Grey curve is the spectrum of post-aragonite CaCO$_3$ collected outside of the heated region. Black vertical bars are computed Raman modes of $P2_1/c$-CaCO$_3$ (bottom) and post-aragonite CaCO$_3$ (top) corrected upwards in frequency by 1.5 % and 0.5 %, respectively. Height of the bars is proportional to the band intensity. The peak indicated by question mark deviates significantly from the $P2_1/c$-CaCO$_3$ model and may be due to the unheated CaCO$_3$, its yet unidentified phase, or minor non-molecular CO$_2$ formed upon CaCO$_3$ thermal decomposition on Pt chunks. (**B**) Experimental spectrum of CaCO$_3$ laser-heated at 105 GPa in comparison with the theoretical spectra of $P2_1/c$ and $C222_1$-CaCO$_3$ at 105 GPa as computed by LDA-DFT. Grey areas are guides to compare the computed spectra with experiment.

Furthermore, we reproduced the frequencies and intensities of all experimentally observed new Raman bands in our LDA-DFT computations of the Raman spectrum of $P2_1/c$-CaCO$_3$ at 105 GPa (Supplementary Table S2). Please note that our computations systematically yielded ~ 1.5 % lower frequencies for all corresponding Raman bands observed in experiment, but when corrected for that, show a remarkable agreement with the experimental spectrum (Fig. 4). Such correction is justified because LDA-DFT yields an equilibrium volume that deviates from experimental observations by up to a few percent (*e.g.* [33]). In addition, we computed a Raman spectrum of $C222_1$-CaCO$_3$ at 105 GPa (Supplementary Table S3), which, expectedly, shows a C-O vibron frequency (996 cm$^{-1}$) that is very close to that in $P2_1/c$-CaCO$_3$ (1011 cm$^{-1}$). Despite this similarity, Raman bands in the 600-850 cm$^{-1}$ spectral range show subtle yet important differences between the $C222_1$ and $P2_1/c$ structures. This difference is likely due to the contrasting packing of the CO$_4$-chains in the structures, which results in slightly different frequencies of deformation modes in CO$_4$-units. As is clear from Figure 4B, the $P2_1/c$ model gives a better agreement with the experiment than the $C222_1$ structure, providing strong spectroscopic evidence for $P2_1/c$-CaCO$_3$ at 105 GPa.

Upon decompression, we could follow the major Raman bands of $sp^3$-bonded CaCO$_3$ down to 57 GPa (Fig. 5). The pressure-frequency dependence of these bands appears consistent with that computed for $P2_1/c$-CaCO$_3$, in support of the band assignment and product identification. Below 57 GPa, however, we could not observe any Raman bands that can be reliably assigned to $P2_1/c$-CaCO$_3$. Evidently, this indicates a full transformation to an $sp^2$-bonded CaCO$_3$ phase below 57 GPa, as is also recorded in the intensification of the band at ~ 1200 cm$^{-1}$, which is representative of CO$_3$ groups (symmetric stretch). Identification of this phase was outside the scope of this work. We note, however, that the CaCO$_3$ system is rich in metastable phases (*e.g.* [16]) and it is possible that the CaCO$_3$ phase formed on unloading to 45 GPa is different from post-aragonite.

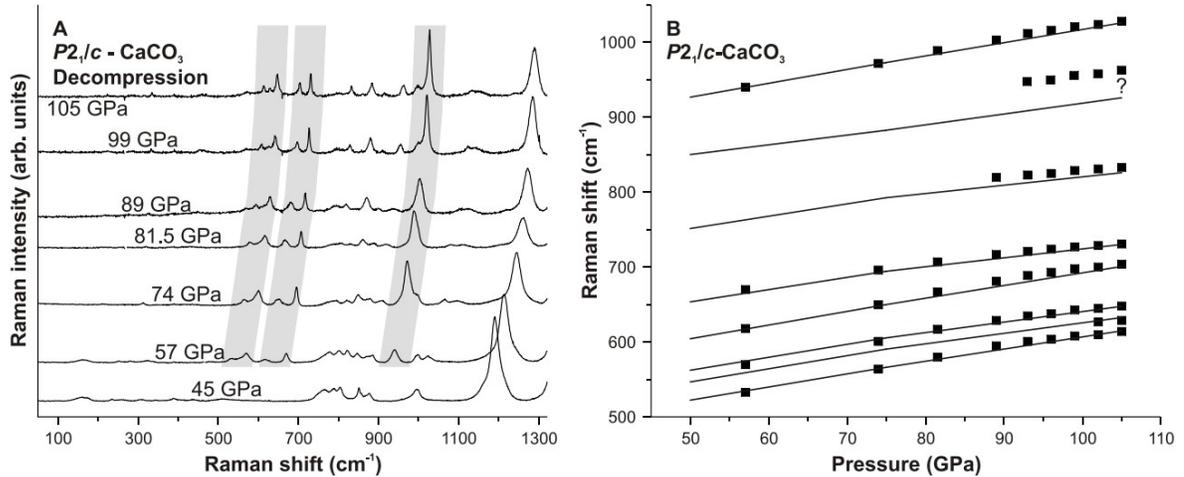

**Figure 5.** (A) Raman spectra (λ = 488 nm) of CaCO$_3$ collected on decompression after laser heating at 105 GPa. Grey areas show characteristic Raman bands of $P2_1/c$-CaCO$_3$. (B) Pressure dependencies of the experimentally observed (squares) and computed (lines) Raman bands of $P2_1/c$-CaCO$_3$ (frequency is corrected upwards by 1.5 %). $P2_1/c$-CaCO$_3$ is preserved down to P = 57 GPa. The error bar for experimental measurements is not shown as it is smaller than the symbols (black squares).

### Mechanism of the $sp^2$-$sp^3$ transition in CaCO$_3$

To reveal the mechanism of the $Pmmn$-CaCO$_3$ → $P2_1/c$-CaCO$_3$ structural phase transition we performed variable-cell nudged elastic band (VCNEB) [34] simulations at 100 GPa, as implemented in the USPEX code [35,36]. First, we obtained an initial trajectory between the two phases using a new algorithm developed by Graf and Stevanovic (publication in preparation) to map crystal structures onto each other. The mapping algorithm relies on criteria of minimizing the total Euclidian distance between the corresponding atoms in the end structures and minimizing the change in their coordination along the map (pathway). The initial pathway was subsequently refined by the VCNEB method for the minimum-energy pathway. Both $Pmmn$ → $P2_1/c$ and $P2_1/c$ → $Pmmn$ paths were prepared and then optimized with VCNEB (in general, this algorithm may find different paths for forward and backward transitions), and the lowest-enthalpy path is presented in detail here. VCNEB calculations required forces and stresses, which were computed by VASP [37] at the PBE-GGA level of theory [38]. Our VCNEB calculations started with 10 intermediate images, and this number automatically increased whenever the path became longer. Climbing image – descending image technique [39] was used to precisely locate transition states (TS) and intermediate minima (corresponding to potential metastable intermediate phases (IP)). Spring constants varied from 3 to 6 eV/Å$^2$. VCNEB calculations were run for 1000 steps, enabling accurate and well-converging results. At the pressure of 100 GPa, $P2_1/c$-CaCO$_3$ phase is more stable by 0.02 eV/atom than post-aragonite. The barrier height is quite large, 0.14 eV/atom (or 0.70 eV/f.u.), implying that this transition is kinetically feasible only at high temperatures, in agreement with experimental results of this work.

One important distinction between the crystal structures of $sp^2$- and $sp^3$-CaCO$_3$ is that CO$_3$ groups in post-aragonite are isolated while CO$_4$ groups in $P2_1/c$-CaCO$_3$ are corner-linked into pyroxene-like chains. Accordingly, the transformation mechanism is quite complex and can be divided into four stages (Fig. 6): each stage corresponds to an energy minimum, and boundaries between them correspond to transition states (TS). In the first stage of the transformation, the post-aragonite structure distorts gradually with all CO$_3$-triangles becoming non-planar. This distortion becomes critical at transition state #1 (TS$_1$) triggering the second stage of the transition with all carbon atoms forming additional bonds with oxygen atoms of the next layer, stitching isolated CO$_3$-groups into infinite chains of CO$_4$-tetrahedra. This topology corresponds to a local enthalpy minimum and has a $P2_1$ symmetry (intermediate phase #1, IP$_1$). However, the enthalpy minimum of IP$_1$ is very shallow (Fig. 6). Towards the transition state TS$_2$, one of the C-O bonds of the original CO$_3$-triangle gradually elongates and eventually breaks. In the third stage, between the transition states TS$_2$ and TS$_3$, yet another metastable structure with a $P2_1$ symmetry appears, featuring flat and non-coplanar CO$_3$-triangles and a shallow enthalpy minimum. As this structure distorts towards the transition state TS$_3$, carbonate triangles reorient, nearby oxygens displace towards them, and eventually one obtains infinite chains of CO$_4$-tetrahedra in the same topology as in the $P2_1/c$ structure. The final, fourth, stage of the transformation is just a relaxation towards the theoretically predicted $P2_1/c$-CaCO$_3$ structure [12].

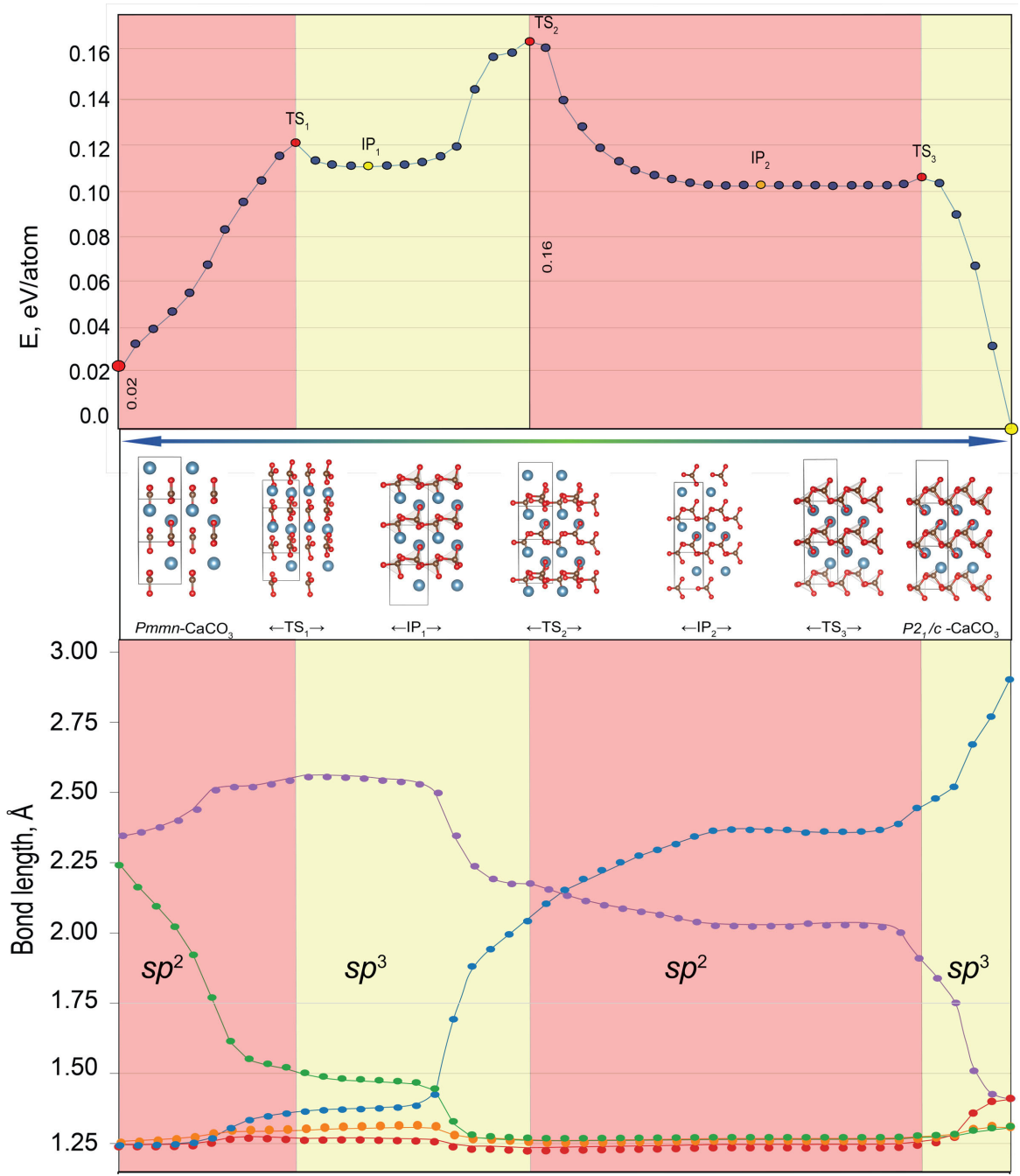

**Figure 6.** Mechanism of the *Pmmn* (post-aragonite) → *P2$_1$/c* transition of CaCO$_3$ at 100 GPa. Structures of initial post-aragonite phase, transition states TS$_1$, TS$_2$ and TS$_3$, intermediate phases IP$_1$, IP$_2$ and final *P2$_1$/c* of CaCO$_3$ are shown (for clarity, we highlighted CO$_4$-tetrahedra). The evolution of five shortest C-O distances is shown across the proposed transition path.

Transition states define the crossover between different topologies – *i.e.* the point at which chemical bonds are formed or broken. It is very tempting to think of some maximum bond lengths characteristic of a given pair of atoms (*e.g.* C-O), beyond which bonds break. However, our results show this not to be the case as the values of critical C-O bond lengths vary for different transitions. This suggests that the phase transitions are driven not just by the nearest-neighbor interactions, but longer-range interactions and cooperative effects are important.

Three fundamental comments are due regarding the mechanism of this phase transition. First, the intermediate minima ($IP_1$ and $IP_2$) in this case are so shallow that they are unlikely to be quenched in the experiment: these minima are not strongly kinetically protected and will rapidly decay into post-aragonite or $P2_1/c$, respectively. The role of these intermediate minima is to be "stepping stones" on the transition pathway, lowering the overall barrier. This is in contrast with the case of BH, a newly predicted compound, where the phase transition involved a very deep and most likely experimentally obtainable, intermediate phase [40]. Second, the transition mechanism discussed here is the best mechanism that we could find (*i.e.* with the lowest activation enthalpy). However, as we did not perform an exhaustive search over transition paths, we cannot rule out the possibility of other mechanisms. At the moment, there is no algorithm for predicting globally optimal transition pathway, even within the mean-field picture. Third, the mechanism we just presented is based on the mean-field picture with all unit cells undergoing the same evolution at a given time. In reality, phase transitions occur via nucleation and growth; thus, the mean-field approach access crude but crystallographically and intuitively tractable models. Full exploration of nucleation and growth phenomena requires very large systems (with $10^2$-$10^4$ atoms) and advanced sampling techniques, such as transition path sampling (*e.g.* [41]); we refer the reader to our recent works employing this methodology (also implemented in the USPEX code) [42,43] and note that such simulations require an accurate forcefield and at the *ab initio* level of theory are computationally unaffordable at the moment.

**Experimental evidence for $sp^3$-bonded carbonates**

Identification of $sp^3$-bonded carbonates solely based on XRD is problematic as it requires precise structure determination, which is often challenging at high pressure. Most previous reports on $sp^3$-carbonates in $MgCO_3$ and $FeCO_3$ systems relied on LeBail-type fits of theoretically predicted structures to experimentally observed power-like XRD patterns. For example, Ref. [44] have reported $sp^3$-$MgCO_3$ at $P \sim 80$ GPa and $T \sim 2000$ K based on the match of XRD to the theoretical prediction of Ref. [11]. One notable exception is the report of $Mg_2Fe_2C_4O_{13}$ with tetrahedrally-coordinated carbon at 135 GPa [45] with single-crystal structure solution methods applied to a multi-grain sample synthesized in the $(Mg,Fe)CO_3$ system. We summarize previous experimental reports on $sp^3$-carbonates in the Table 1.

**Table 1.** Summary of experimental reports on carbonates with tetrahedrally-coordinated carbon.

| References | System | Space group | P, GPa | Problems |
|---|---|---|---|---|
| Ref. [19] | $CaCO_3$ | $C222_1$ | 130 | No spectroscopic probe for $sp^3$-carbon |
| Ref. [44] | $MgCO_3$ | $C2/m$ and | 82 | LeBail fit, no spectroscopic probe for $sp^3$-carbon |

| | | $P2_1/a$ | | |
|---|---|---|---|---|
| Ref. [46] | FeCO$_3$ | - | 80 | LeBail fit; no spectroscopic probe for $sp^3$-carbon |
| Ref. [47] | (Mg,Fe)CO$_3$ | $C2/m$ and $P2_1/a$ | 80 | LeBail fit |
| Ref. [45] | (Mg,Fe)CO$_3$ | $C2/c$ | 135 | No spectroscopic probe for $sp^3$-carbon |

Unlike XRD, vibrational spectroscopy provides bonding fingerprints of the material and is particularly sensitive to the carbon hybridization and chemical environment (*e.g.* [48]). As such, Raman and/or infrared spectroscopy provide independent evidence for tetrahedrally-coordinated carbon and must be used together with crystallographic probes for a reliable identification of $sp^3$-carbonates in high-pressure experiments. Realizing weaknesses of XRD probes in identifying the bond character, Boulard et al., [47] reported on synchrotron infrared absorption experiments in (Mg$_{0.25}$Fe$_{0.75}$)CO$_3$ at 80 GPa, noting a new band that is characteristic of the C-O asymmetric stretching vibration in CO$_4$-groups. The band assignment relied on first-principles calculations of the infrared spectrum of $sp^3$-MgCO$_3$ ($P2_1/a$ space group). However, other theoretically predicted bands were not fully assigned in the experiment [47].

In contrast to previous studies, here we provided strong spectroscopic evidence of $sp^3$-carbonates. Specifically, the intense Raman band at ~ 1025 cm$^{-1}$ (at 105 GPa) and its pressure dependence (~ 1.8 cm$^{-1}$/GPa) in $P2_1/c$-CaCO$_3$ are characteristic of the symmetrical stretching vibration in its CO$_4$-groups. In principle, these spectroscopic features can be used in future studies of $sp^3$-carbonates at high pressure to confirm fourfold carbon coordination.

Our results are also important to validate density functionals used in crystal structure predictions. Pickard and Needs [12] noted that Perdew-Burke-Ernzerhof generalized gradient approximation (PBE-GGA) and local density approximation (LDA) yield essentially similar transition pressures, and thus, provide accurate description of the electronic structures. Here we identified the $sp^3$-CaCO$_3$ phase and the $sp^2$-$sp^3$ crossover pressure (105 GPa), which appears to be ~ 30 GPa higher than the theoretically predicted transition pressure of 76 GPa (at 0 K), suggesting that the entropy term in the free energy is substantial. We showed that high temperature is required to overcome the kinetic barriers associated with the $sp^2$-$sp^3$ transition, indicating that complex energy landscapes are typical not only of pure carbon but of carbonates just as well. As a result, a variety of metastable $sp^2$-CaCO$_3$ polymorphs have been observed at P < 40 GPa [16]. Results of this study suggest that $sp^3$-CaCO$_3$ may also have a number of metastable structures accessible through compression without high-T annealing. In this regard, the Raman signature of $sp^3$-carbonates may come in useful to diagnose tetrahedrally-coordinated carbon.

**Geochemical and geophysical implications of *sp*$^3$-carbonates in the lowermost mantle**

The equilibrium composition of mantle carbonates is governed by the chemical reactions with surrounding minerals [13,14,49] and thermodynamic stability of corresponding carbonate phases. Due to the chemical interaction with pyroxene or bridgmanite in the mantle, $CaCO_3$ transforms to Fe-bearing magnesite (up to 10 % Fe [50]) at 2-80 GPa [51-54] despite several phase transitions in *sp*$^2$-$CaCO_3$ which can modify the chemical equilibrium in this pressure range [9,12]. Also, the spin transition in Fe-bearing $MgCO_3$ at P ~ 45 GPa may promote iron solubility in the carbonate phase due to crystal field effects [55] and ionic size similarity of low spin $Fe^{2+}$ with $Mg^{2+}$ [56], but this has never been quantitatively addressed in experiment. The *sp*$^2$-*sp*$^3$ transition in $MgCO_3$ at P ~ 80 GPa further upholds the Mg-rich carbonate composition, as revealed by a computation of enthalpies in the reaction $MgCO_3 + CaSiO_3 = CaCO_3 + MgSiO_3$ as a function of pressure and accounting for phase transitions [11,12]. The theoretically predicted *sp*$^2$-*sp*$^3$ transition in $CaCO_3$ at 76 GPa eventually stabilizes $CaCO_3$ against $MgCO_3$ at P > ~100 GPa [12]. Here we have synthesized the predicted $P2_1/c$-$CaCO_3$ at P ~ 105 GPa and T ~ 2000 K, about 30 GPa higher than the theoretically predicted *sp*$^2$-*sp*$^3$ transition pressure at 0 K. Taking into account this 30 GPa discrepancy we propose that the crossover to Ca-carbonates in Earth (*i.e.* at high temperature) may be expected at P ~ 135 GPa, which corresponds to the pressure at the core-mantle boundary. This inference can be tested via high-pressure studies of chemical reactions in mechanical mixtures of $MgCO_3$ with $CaSiO_3$ or $CaCO_3$ with $MgSiO_3$ at high pressure and temperature.

**Conclusions**

In summary, we located the *sp*$^2$-*sp*$^3$ transition in $CaCO_3$ and identified the $P2_1/c$-$CaCO_3$ at P > 105 GPa using x-ray diffraction and Raman spectroscopy. Using first-principles methods, we showed that the mechanism of the *sp*$^2$-*sp*$^3$ crossover in $CaCO_3$ involves several intermediate phases with *sp*$^2$ and *sp*$^3$ bonding motifs. Finally, our results support the idea of the crossover in the carbonate crystal chemistry that leads to Ca-rich carbonates at the base of the mantle.

**Acknowledgments**

This work was supported by the National Science foundation, grants NSF EAR/IF 1531583, NSF EAR-1520648 and NSF EAR/IF-1128867, the Army Research Office (56122-CH-H), the Carnegie Institution of Washington and Deep Carbon Observatory. S.S.L. was partly supported by state assignment project No. 0330-2014-0013. Portions of this work were performed at GeoSoilEnviroCARS (The University of Chicago, Sector 13), Advanced Photon Source (APS), Argonne National Laboratory. GeoSoilEnviroCARS is supported by the National

Science Foundation - Earth Sciences (EAR-1128799) and Department of Energy- GeoSciences (DE-FG02-94ER14466). This research used resources of the Advanced Photon Source, a U.S. Department of Energy (DOE) Office of Science User Facility operated for the DOE Office of Science by Argonne National Laboratory under Contract No. DE-AC02-06CH11357. We thank Dr. E. V. Alexandrov for his assistance in topological analysis. P.N.G. was supported by the Ministry of Education and Science of Russian Federation (No 14.B25.31.0032 and MK-3417.2017.5). A. F. G. was partly supported by the Chinese Academy of Sciences visiting professorship for senior international scientists (Grant No. 2011T2J20), Recruitment Program of Foreign Expert, the National Natural Science Foundation of China (grant number 21473211), and the Chinese Academy of Sciences (grant number YZ201524). A.R.O was supported by Russian Science Foundation (grant 16-13-10459).

# SUPPLEMENTARY INFORMATION

## Supplementary Figures

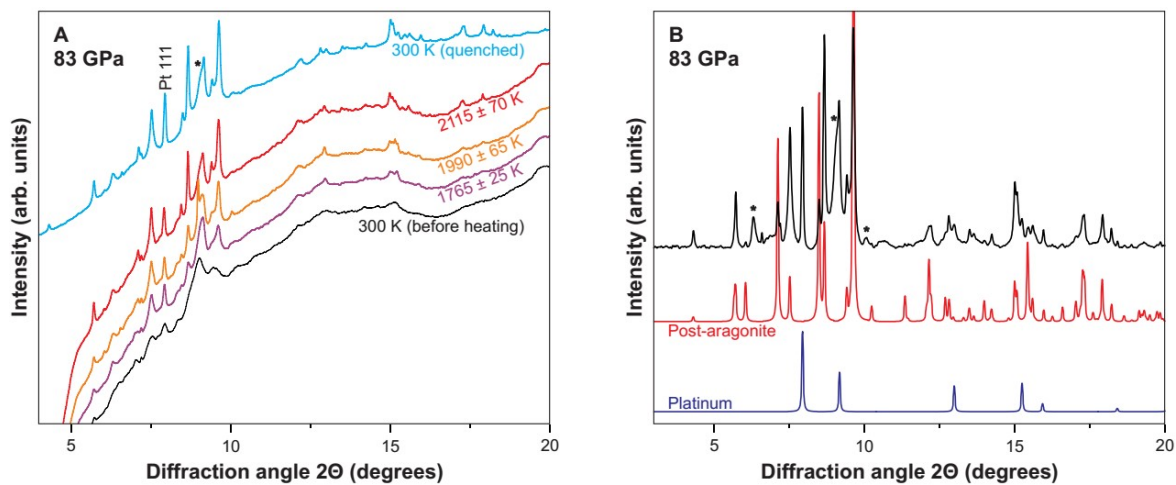

**Supplementary Figure S1.** (**A**) X-ray diffraction (XRD) of CaCO$_3$ before, at *T*, and after heating at 83 GPa (with background). (**B**) Background-subtracted XRD pattern (black line) of the heated CaCO$_3$ at 83 GPa after cooling to 300 K. Red and blue lines are powder XRD patterns of post-aragonite CaCO$_3$ and Pt, respectively. Asterisks and black boxes mark some of the diffuse peaks of remnant CaCO$_3$. X-ray energy is 42 keV.

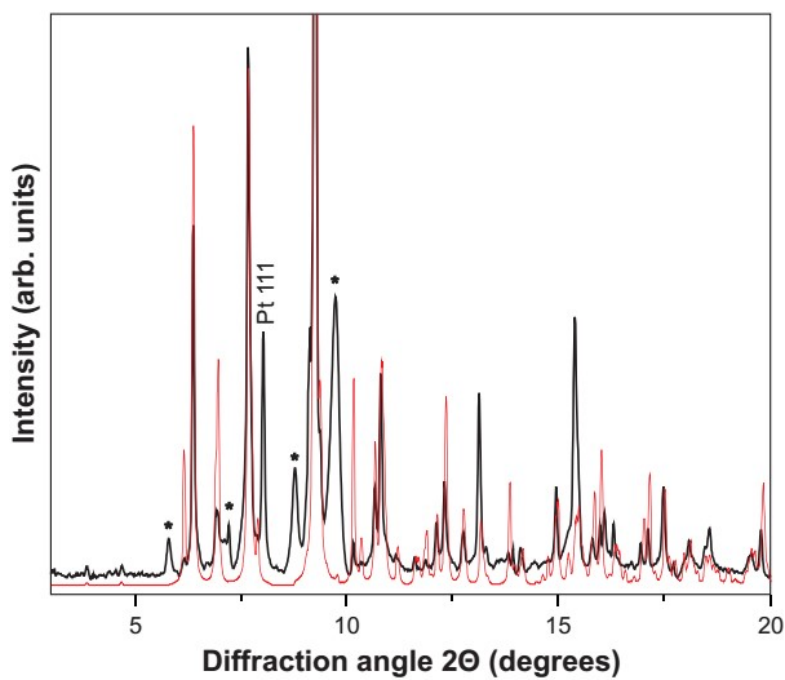

**Supplementary Figure S2.** Room-temperature x-ray diffraction (XRD) of CaCO$_3$ laser-heated at 105 GPa (black line) in comparison to the powder XRD pattern of the $P2_1/c$-CaCO$_3$ model [12] (red line). Asterisks and black boxes mark some of the diffuse peaks of remnant CaCO$_3$. X-ray energy is 42 keV.

# Supplementary Tables

**Supplementary Table S1.** Lattice parameters of the $P2_1/c$-, $C222_1$-, and $Pmmn$-$CaCO_3$ models as refined from the experimental XRD data at 105 GPa.

| Model / Lattice parameter | a, Å | b, Å | c, Å | β, degrees |
|---|---|---|---|---|
| $P2_1/c$-$CaCO_3$ | 4.5288(13) | 3.3345(3) | 9.0927 (24) | 105.57(9) |
| $C222_1$-$CaCO_3$ | 5.5057(4) | 7.2389(3) | 3.3425(2) | - |
| $Pmmn$-$CaCO_3$ | 3.9360(6) | 4.4372(3) | 3.9049(4) | - |

**Supplementary Table S2.** Vibrational properties of $P2_1/c$-$CaCO_3$ at 105 GPa as computed by LDA-DFT.

| Mode # | Frequency (cm$^{-1}$) | Frequency (THz) | IR intensity | Raman intensity | Depolarization factor |
|---|---|---|---|---|---|
| 1 | 0 | 0 | 0 | 0 | 0.2089 |
| 2 | 0 | 0 | 0 | 0 | 0.4892 |
| 3 | 0 | 0 | 0 | 0 | 0.2933 |
| 4 | 173.5 | 5.2014 | 0 | 0.0201 | 0.75 |
| 5 | 208.52 | 6.2513 | 0.1822 | 0 | 0.7023 |
| 6 | 218.25 | 6.5431 | 0 | 1.033 | 0.7203 |
| 7 | 273.91 | 8.2116 | 0.4094 | 0 | 0.232 |
| 8 | 312.96 | 9.3823 | 0 | 0.3429 | 0.2483 |
| 9 | 313.87 | 9.4095 | 0 | 0.0932 | 0.75 |
| 10 | 328.24 | 9.8403 | 0 | 0.2048 | 0.0856 |
| 11 | 350.07 | 10.4947 | 0 | 0.0345 | 0.75 |
| 12 | 370.5 | 11.1073 | 0.0593 | 0 | 0.6904 |
| 13 | 379.49 | 11.3767 | 0 | 0.1724 | 0.7051 |
| 14 | 435.31 | 13.0502 | 0 | 0.1706 | 0.75 |
| 15 | 441.23 | 13.2278 | 20.9243 | 0 | 0.7494 |
| 16 | 441.67 | 13.241 | 0.0312 | 0 | 0.7499 |
| 17 | 449.89 | 13.4875 | 0 | 1.9539 | 0.7496 |
| 18 | 482.54 | 14.4661 | 0 | 0.1497 | 0.75 |
| 19 | 484.64 | 14.5292 | 11.2104 | 0 | 0.6888 |
| 20 | 494.59 | 14.8276 | 3.9715 | 0 | 0.612 |
| 21 | 505.47 | 15.1535 | 0 | 0.9328 | 0.75 |
| 22 | 514.42 | 15.4218 | 10.1563 | 0 | 0.7457 |
| 23 | 570.83 | 17.1132 | 0 | 1.2989 | 0.3694 |
| 24 | 576.9 | 17.2951 | 0.0496 | 0 | 0.7068 |
| 25 | 589.19 | 17.6636 | 9.0206 | 0 | 0.6326 |
| 26 | 605.96 | 18.1662 | 0 | 10.0644 | 0.75 |
| 27 | 623.51 | 18.6923 | 0 | 5.5571 | 0.2656 |
| 28 | 623.74 | 18.6991 | 6.3122 | 0 | 0.2648 |
| 29 | 638.35 | 19.1374 | 0 | 31.0868 | 0.75 |
| 30 | 652.25 | 19.554 | 0.2032 | 0 | 0.7494 |
| 31 | 656.42 | 19.679 | 0.0026 | 0 | 0.6994 |
| 32 | 690.65 | 20.7052 | 0 | 14.3927 | 0.3381 |
| 33 | 718.39 | 21.5367 | 0.2715 | 0 | 0.7481 |
| 34 | 719.63 | 21.574 | 0 | 20.1884 | 0.7439 |
| 35 | 779.3 | 23.3628 | 0 | 0.7965 | 0.0637 |
| 36 | 784.68 | 23.5243 | 63.689 | 0 | 0.5959 |
| 37 | 797.73 | 23.9155 | 9.5441 | 0 | 0.7248 |
| 38 | 814.05 | 24.4045 | 0 | 14.9039 | 0.75 |
| 39 | 840.02 | 25.1833 | 1.5425 | 0 | 0.7233 |

| Mode # | Frequency (cm⁻¹) | Frequency (THz) | IR intensity | Raman intensity | Depolarization factor |
|---|---|---|---|---|---|
| 40 | 847.03 | 25.3932 | 0 | 1.1494 | 0.75 |
| 41 | 863.18 | 25.8775 | 0 | 8.0109 | 0.75 |
| 42 | 890.86 | 26.7074 | 0.6486 | 0 | 0.5888 |
| 43 | 895.47 | 26.8454 | 0 | 7.2391 | 0.6332 |
| 44 | 897.97 | 26.9205 | 21.9671 | 0 | 0.6424 |
| 45 | 905.2 | 27.1371 | 0.0535 | 0 | 0.3265 |
| 46 | 912.41 | 27.3534 | 0 | 17.7336 | 0.1821 |
| 47 | 1000.57 | 29.9963 | 2.546 | 0 | 0.0103 |
| 48 | 1011.05 | 30.3105 | 0 | 127.1435 | 0.0079 |
| 49 | 1131.01 | 33.9068 | 0 | 2.7042 | 0.75 |
| 50 | 1138.17 | 34.1214 | 26.799 | 0 | 0.195 |
| 51 | 1202.25 | 36.0425 | 0 | 4.02 | 0.75 |
| 52 | 1223.71 | 36.6859 | 20.4645 | 0 | 0.7365 |
| 53 | 1332.78 | 39.9558 | 83.6266 | 0 | 0.7498 |
| 54 | 1332.91 | 39.9597 | 0 | 11.085 | 0.75 |
| 55 | 1409.66 | 42.2605 | 6.1204 | 0 | 0.1223 |
| 56 | 1413.35 | 42.3713 | 0 | 10.0924 | 0.1273 |
| 57 | 1435.07 | 43.0223 | 0 | 9.4744 | 0.5871 |
| 58 | 1483.68 | 44.4796 | 31.1199 | 0 | 0.6671 |
| 59 | 1514.84 | 45.4137 | 0 | 6.546 | 0.75 |
| 60 | 1553.62 | 46.5763 | 77.3633 | 0 | 0.6603 |

**Supplementary Table S3.** Vibrational properties of $C222_1$-CaCO$_3$ at 105 GPa as computed by LDA-DFT.

| Mode # | Frequency (cm⁻¹) | Frequency (THz) | IR intensity | Raman intensity | Depolarization factor |
|---|---|---|---|---|---|
| 1 | 0 | 0 | 0 | 0 | 0.75 |
| 2 | 0 | 0 | 0 | 0 | 0.75 |
| 3 | 0 | 0 | 0 | 0 | 0.75 |
| 4 | 210.03 | 6.2965 | 0.0381 | 0.0984 | 0.75 |
| 5 | 218.23 | 6.5422 | 0.3481 | 0.897 | 0.75 |
| 6 | 240.1 | 7.1981 | 0.0003 | 0.0009 | 0.75 |
| 7 | 283.81 | 8.5085 | 0.0003 | 0 | 0.75 |
| 8 | 300.14 | 8.9979 | 0.0011 | 0.0001 | 0.75 |
| 9 | 332.6 | 9.971 | 0.0067 | 0.0004 | 0.75 |
| 10 | 332.76 | 9.9758 | 0 | 0.2009 | 0.3151 |
| 11 | 337.47 | 10.117 | 4.9077 | 0.2023 | 0.75 |
| 12 | 384.49 | 11.5267 | 0.0002 | 0 | 0.75 |
| 13 | 426.34 | 12.7813 | 0.003 | 0.007 | 0.75 |
| 14 | 440.45 | 13.2045 | 0.2366 | 1.3334 | 0.75 |
| 15 | 440.88 | 13.2173 | 0 | 0 | 0.0173 |
| 16 | 447.78 | 13.424 | 0.0002 | 0 | 0.75 |
| 17 | 451.93 | 13.5484 | 26.8854 | 1.4425 | 0.75 |
| 18 | 492.49 | 14.7646 | 0.007 | 0.002 | 0.75 |
| 19 | 493.73 | 14.8016 | 1.6703 | 1.2386 | 0.75 |
| 20 | 494.17 | 14.8149 | 24.4814 | 0.0016 | 0.75 |
| 21 | 505.86 | 15.1653 | 4.21 | 3.3619 | 0.75 |
| 22 | 551.35 | 16.5291 | 0.0016 | 0.004 | 0.75 |
| 23 | 573.24 | 17.1853 | 1.0609 | 1.4055 | 0.75 |
| 24 | 591.76 | 17.7407 | 0 | 0.0077 | 0.75 |
| 25 | 614.79 | 18.431 | 0 | 19.6013 | 0.1583 |
| 26 | 619.09 | 18.5598 | 0.3029 | 32.9517 | 0.75 |

| 27 | 637.15 | 19.1011 | 0 | 0.0461 | 0.1472 |
| 28 | 647.29 | 19.4053 | 0.0003 | 0.0009 | 0.75 |
| 29 | 664.14 | 19.9103 | 0.2501 | 0.0001 | 0.75 |
| 30 | 677.8 | 20.3198 | 0.002 | 0.0058 | 0.75 |
| 31 | 690.38 | 20.6971 | 0 | 8.9335 | 0.426 |
| 32 | 707.62 | 21.2139 | 0 | 0.0006 | 0.6001 |
| 33 | 711.42 | 21.328 | 0.0307 | 18.7983 | 0.75 |
| 34 | 731.82 | 21.9395 | 0 | 0.0016 | 0.75 |
| 35 | 770.74 | 23.1062 | 0 | 0 | 0.75 |
| 36 | 786.02 | 23.5643 | 11.699 | 25.9228 | 0.75 |
| 37 | 806.61 | 24.1816 | 0.0186 | 0.0258 | 0.75 |
| 38 | 849.79 | 25.476 | 2.8018 | 9.0094 | 0.75 |
| 39 | 854.1 | 25.6051 | 0.0003 | 0.0029 | 0.75 |
| 40 | 871.31 | 26.1212 | 0.0174 | 0.0021 | 0.75 |
| 41 | 871.95 | 26.1403 | 0.0003 | 0.0038 | 0.75 |
| 42 | 881.15 | 26.4161 | 1.1213 | 5.6692 | 0.75 |
| 43 | 890.97 | 26.7107 | 10.9965 | 0.0472 | 0.75 |
| 44 | 894.9 | 26.8286 | 0 | 14.5087 | 0.2294 |
| 45 | 921.24 | 27.6181 | 0 | 0.0121 | 0.1111 |
| 46 | 968.82 | 29.0443 | 0 | 0.2781 | 0.0056 |
| 47 | 983.24 | 29.4767 | 62.6936 | 2.8823 | 0.75 |
| 48 | 995.72 | 29.851 | 0 | 140.0152 | 0.0075 |
| 49 | 1068.34 | 32.0279 | 50.9045 | 1.9079 | 0.75 |
| 50 | 1086.84 | 32.5827 | 0.0089 | 0.0006 | 0.75 |
| 51 | 1171.28 | 35.1141 | 0.0024 | 0.0004 | 0.75 |
| 52 | 1196.96 | 35.884 | 18.9827 | 4.0087 | 0.75 |
| 53 | 1225.34 | 36.7348 | 51.9147 | 3.6436 | 0.75 |
| 54 | 1245.1 | 37.3273 | 0.0073 | 0.0007 | 0.75 |
| 55 | 1360.03 | 40.7727 | 0.0007 | 0 | 0.75 |
| 56 | 1384.76 | 41.5141 | 0 | 7.461 | 0.0958 |
| 57 | 1416.35 | 42.4612 | 0 | 0.0002 | 0.0266 |
| 58 | 1459.06 | 43.7415 | 60.2865 | 0.1559 | 0.75 |
| 59 | 1480.42 | 44.3819 | 73.8631 | 6.3651 | 0.75 |
| 60 | 1560.12 | 46.7711 | 0.0009 | 0 | 0.75 |

**Supplementary Table S4.** Vibrational properties of *Pmmn*-CaCO$_3$ (post-aragonite) at 105 GPa as computed by LDA-DFT.

| Mode # | Frequency (cm$^{-1}$) | Frequency (THz) | IR intensity | Raman intensity | Depolarization factor |
| --- | --- | --- | --- | --- | --- |
| 1 | 0 | 0 | 0 | 0 | 0.3095 |
| 2 | 0 | 0 | 0 | 0 | 0.75 |
| 3 | 0 | 0 | 0 | 0 | 0.7498 |
| 4 | 239.85 | 7.1904 | 0 | 0.5682 | 0.75 |
| 5 | 283.97 | 8.5133 | 0.0002 | 1.5292 | 0.75 |
| 6 | 297.49 | 8.9184 | 0 | 0.0017 | 0.75 |
| 7 | 346.56 | 10.3895 | 0 | 4.0606 | 0.3295 |
| 8 | 394.89 | 11.8385 | 7.3676 | 0.0004 | 0.75 |
| 9 | 395.67 | 11.8619 | 15.3288 | 0.0024 | 0.75 |
| 10 | 397.69 | 11.9225 | 0.0005 | 0.2161 | 0.75 |
| 11 | 437.17 | 13.106 | 0 | 2.5285 | 0.3858 |
| 12 | 471.65 | 14.1398 | 0.0416 | 5.4371 | 0.75 |
| 13 | 529.91 | 15.8862 | 4.9378 | 0.0025 | 0.75 |
| 14 | 585.31 | 17.5472 | 1.101 | 0.0543 | 0.75 |
| 15 | 602.08 | 18.0498 | 12.7315 | 0.0001 | 0.0186 |
| 16 | 630.88 | 18.9133 | 0.0013 | 0.3585 | 0.75 |
| 17 | 659.19 | 19.7621 | 0 | 7.7721 | 0.75 |
| 18 | 719.49 | 21.5697 | 0.0015 | 0.0148 | 0.75 |
| 19 | 767.97 | 23.0232 | 0.3733 | 13.1236 | 0.75 |
| 20 | 775.94 | 23.262 | 9.113 | 0.5664 | 0.75 |
| 21 | 803.67 | 24.0933 | 9.9435 | 0.0095 | 0.75 |
| 22 | 813.64 | 24.3923 | 0.492 | 0.1354 | 0.75 |
| 23 | 862.85 | 25.8677 | 0.002 | 29.4427 | 0.4727 |
| 24 | 870.63 | 26.1009 | 2.6172 | 0.0299 | 0.4034 |
| 25 | 1281.13 | 38.4072 | 0.0018 | 350.143 | 0.0192 |
| 26 | 1291.31 | 38.7125 | 0.2826 | 1.5192 | 0.0179 |
| 27 | 1652.68 | 49.546 | 71.9921 | 0 | 0.75 |
| 28 | 1664.4 | 49.8973 | 0.0002 | 14.2517 | 0.7026 |
| 29 | 1738.5 | 52.1191 | 0 | 0.5207 | 0.75 |
| 30 | 1840.75 | 55.1843 | 61.7667 | 0.0002 | 0.0952 |